# Status of the Stratospheric Observatory for Infrared Astronomy (SOFIA)

R. D. Gehrz[a], E. E. Becklin[b], J. de Buizer[b], T. Herter[c], L. D. Keller[d], A. Krabbe[e], P. M. Marcum[f], T. L. Roellig[f], G. H. L. Sandell[b], P. Temi[f], W. D. Vacca[b], E. T. Young[b], and H. Zinnecker[e,g]

[a]Department of Astronomy, School of Physics and Astronomy, 116 Church Street, S. E., University of Minnesota, Minneapolis, MN 55455, USA
[b]Universities Space Research Association, NASA Ames Research Center, MS 211-3, Moffett Field, CA 94035, USA
[c]Astronomy Department, 202 Space Sciences Building, Cornell University, Ithaca, NY 14853-6801, USA
[d]Department of Physics, Ithaca College, Ithaca, NY 14850, USA
[e]Deutsches SOFIA Institut, Universität Stuttgart, Pfaffenwaldring 31, D-70569 Stuttgart, Germany
[f]NASA Ames Research Center, MS 245-6, Moffett Field, CA 94035, USA
[g]SOFIA Science Center, NASA Ames Research Center, MS N211-3, Moffett Field, CA 94035, USA

## Abstract

The Stratospheric Observatory for Infrared Astronomy (SOFIA), a joint U.S./German project, is a 2.5-meter infrared airborne telescope carried by a Boeing 747-SP that flies in the stratosphere at altitudes as high as 45,000 feet (13.72 km). This facility is capable of observing from 0.3 μm to 1.6 mm with an average transmission greater than 80 percent. SOFIA will be staged out of the NASA Dryden Flight Research Center aircraft operations facility at Palmdale, CA. The SOFIA Science Mission Operations (SMO) will be located at NASA Ames Research Center, Moffett Field, CA. First science flights began in 2010 and a full operations schedule of up to one hundred 8 to 10 hour flights per year will be reached by 2014. The observatory is expected to operate until the mid 2030's. SOFIA's initial complement of seven focal plane instruments includes broadband imagers, moderate-resolution spectrographs that will resolve broad features due to dust and large molecules, and high-resolution spectrometers capable of studying the kinematics of atomic and molecular gas at sub-km/s resolution. We describe the SOFIA facility and outline the opportunities for observations by the general scientific community and for future instrumentation development. The operational characteristics of the SOFIA first-generation instruments are summarized. The status of the flight test program is discussed and we show First Light images obtained at wavelengths from 5.4 to 37 μm with the FORCAST imaging camera. Additional information about SOFIA is available at http://www.sofia.usra.edu and http://www.sofia.usra.edu/Science/docs/SofiaScienceVision051809-1.pdf

# 1. Program Overview

SOFIA, a joint project of NASA and the German Space Agency (DLR), is a 2.5-meter telescope in a Boeing 747SP aircraft (Figure 1) designed to make sensitive IR measurements of a wide range of astronomical objects (Stutzki 2006, Becklin et al. 2007, Gehrz et al. 2009). SOFIA will fly at altitudes up to 45,000 feet (13.72 km), above 99.8% of the atmospheric $H_2O$ vapor. The facility will enable observations at wavelengths from 0.3 µm to 1.6 mm with ≥ 80% transmission. At SOFIA's service altitude, where the typical precipitable atmospheric water ($H_2O$) column depth is less than 10 µm (a hundred times lower than at good terrestrial sites), SOFIA will provide unprecedented spectral and spatial resolution observations in large parts of the spectrum that are completely inaccessible from the ground (Figure 2). Although some strong water absorption lines remain, the pressure broadening is much reduced so that high resolution spectroscopy is possible between these lines. The SOFIA telescope design and its evolving instrument complement build upon the heritage of NASA's Kuiper Airborne Observatory (KAO, Gillespie 1981), a 0.9-meter infrared telescope that flew from 1974-1995 in a Lockheed C141 Starlifter aircraft.

SOFIA is especially well-suited for chemical and dynamical studies of warm material in the universe, for observations of deeply embedded sources, and for rapid response to transient events. Current plans are to operate the SOFIA Observatory until the mid 2030's. It will join the Spitzer Space Telescope (Werner et al. 2004; Gehrz et al. 2007), Herschel Space Observatory (Pilbratt 2003), and James Webb Space Telescope (JWST, Gardner et al. 2006) as one of the primary facilities for observations in the thermal IR and sub-millimeter. In fact, the SOFIA mission will cover a larger range of wavelengths for a longer span of time than any planned space mission (Figure 3). The synergies and complementarities between SOFIA and the other ground-based and space observatories that will be operating during SOFIA's mission lifetime have been discussed in detail by Gehrz and Becklin (2008), Gehrz et al. (2009), and "The Science Vision for the Stratospheric Observatory for Infrared Astronomy" (VISION 2009).

SOFIA will also be a test bed for new technologies and a training ground for a new generation of instrumentalists. It will provide community-wide opportunities for forefront science, invaluable hands-on experience for young researchers, and an extensive and unique education and public outreach program. With observing flexibility and the ability to deploy new and updated instruments, the observatory will make important contributions towards understanding a variety of astrophysical problems well into the 21st century.

*1.1 Science operations, flight operations, and schedule*

First test flights of the observatory began in April 2007 at L-3 Communications in Waco, Texas after which it was ferried to NASA's Dryden Aircraft Operations Facility (DAOF) in Palmdale, CA. Further flight testing and development has been conducted at DAOF, the home base for flight operations of the SOFIA aircraft, *Clipper Lindbergh*. Closed door testing was completed in January, 2008. The first flight of the DAOF open door test series occurred on December 9, 2009, and the door was opened to expose the full aperture of the telescope a week later on December 18, 2009 (Figure 1). Initial imaging and telescope tracking tests were conducted at night on the ground at Palmdale, CA, followed by SOFIA acquiring its first light IR image on May, 26, 2010. Science flights began in fall 2010, and the flight frequency will ramp up steadily until 2014, when SOFIA will begin making more than one hundred 8-10 hour scientific flights per year until the mid 2030's.

SOFIA is funded jointly by NASA and the German Space Agency (DLR), with the two agencies splitting the operational costs and observing time by 80% and 20% respectively. The Universities Space Research Association (USRA) and the Deutsches SOFIA Institut (DSI) in Stuttgart, Germany manage science and mission operations for NASA and DLR. Science support for the SOFIA user community will be provided by the SOFIA Science Mission Operations (SMO) at NASA Ames Research Center, Moffett Field, California and the DSI at the University of Stuttgart, Stuttgart, Germany. In addition to operations out of its home base at DAOF, SOFIA will also operate from other bases world wide, including some in the southern hemisphere.

The SOFIA Program will support up to ~50 science teams per year, selected from peer reviewed proposals. An on-going instrument development program will ensure that the observatory maintains state-of-the-art performance during its lifetime. The next call for new generation instruments will occur in mid-2011. Science operations will start with a phased approach in order to best utilize the platform to guide scientific optimization of the observatory and the development of its next generation of instruments. The Early Short Science Program with the FORCAST and GREAT instruments (described in Section 2 below), selected in response to proposals solicited during 2008, began in late 2010 and will continue in early 2011. These first science flights, limited in scope, will be based on science collaborations between General Investigators (GIs) and the Principal Investigators (PIs) of those instruments in order to ensure early science productivity. Early Basic Science observations will begin in 2011 in response to a call for General Observer (GO) science proposals that was released on April 19, 2010 with a submission deadline of July 30, 2010. Sixty unique proposals were received requesting a total of 234 hours for FORCAST and 42 hours for GREAT. The proposals were reviewed for scientific merit during fall, 2010 and 52.1 and 17.4 hours of time were awarded for FORCAST and GREAT respectively. There will be new science proposal solicitations every 12 months starting in 2011. About 20 GO science flights are planned annually at the start of science operations, with the annual flight rate ramping up steadily until the full annual rate of ~100 per year is achieved in 2014.

*1.2 Special advantages of SOFIA*

As a near-space observatory that comes home after every flight, SOFIA has the great advantage that its science instruments (SIs) can be adjusted, repaired (in flight if necessary), and exchanged regularly as science requirements dictate. New SIs and modifications of old SIs can take advantage of advanced new technologies that are not yet space-qualified. SIs that are too large, massive, complex and sophisticated for space flight as well as SIs instruments having power needs and heat dissipation that are too great for spacecraft can be flown on SOFIA. Unlike ground-based observatories whose nightly operations are seriously compromised or disabled by weather conditions, observing on SOFIA generally occurs above most adverse weather. SOFIA's ability to fly from any airport that can accommodate a 747 is a unique and a key enabling aspect of the mission. The observatory can, in principle, operate from other northern and southern hemisphere bases when necessary to conduct observations at any declination and respond promptly to targets of opportunity in any part of the sky. This capability will make SOFIA an ideal platform for observing transient and location-specific events such as variable stars, comets, occultations, eclipses, novae, and supernovae. Because the SOFIA telescope can point to elevations as low as 20 degrees, observations can be made of important astrophysical events and solar system objects even when they are close enough to the Sun to be unobservable from IR space missions. In addition to providing the astronomical community with new opportunities for observations at the cutting edge of science, the observatory will give students the opportunity to participate in hands-on development of advanced instrumentation. The development of the next generation of instrumentalists is a high priority for astronomical science. SOFIA's accessibility will facilitate a vigorous education and public outreach program that will eventually enable educators to work with the SI teams to expose their students to the excitement of science.

**2.0 The SOFIA observatory and its instrumentation**

We describe below and in Table 1 the top-level design and performance specifications of the SOFIA Observatory Facility, its initial complement of seven instruments and their expected performance, and the SOFIA Project's plans for developing new instrumentation during the life of the SOFIA mission.

*2.1 The SOFIA telescope and observatory*

The SOFIA telescope (Table 1 and Figures 4 and 5), supplied by DLR under an agreement with NASA as the German contribution to the development of SOFIA, is a bent Cassegrain with a 2.7m (2.5m effective aperture) parabolic primary mirror and a 0.35m diameter hyperbolic secondary mirror with a f/19.6 Nasmyth infrared focus fed by a 45° gold coated dichroic mirror. The unvignetted field of view (FOV) is 8 arcminutes. The infrared Nasmyth focus is 300mm behind the instrument mounting flange. The dichroic mirror allows transmitted optical light to be reflected by a second tertiary behind the dichroic to a visible Nasmyth focus where it is fed into an optical focal plane guiding camera system, the Focal Plane Imager (FPI). Two other imaging and guiding cameras, independent of the FPI, are available: the Wide Field Imager (WFI) and the Fine Field Imager (FFI). Both the WFI and the FFI are attached to the front ring of the telescope. These trackers have larger fields of view and less pointing control accuracy than the focal plane camera. The secondary mirror is attached to a chopping mechanism providing unvignetted chop amplitudes of up to ± 4 arcminutes at chop frequencies between 0 and 20 Hz, programmable by either a user-supplied wave-form or by the telescope control electronics. Asymmetric chop amplitudes as large as 10 arcminutes can be accommodated with some vignetting and image degradation.

The telescope, inertially stabilized by electronic fiber optic gyros, resides in an open cavity in the aft section of the aircraft and views the sky through a port-side doorway (Figures 1 and 4). The door, which is opened at flight altitude, has a rigid upper segment and a flexible lower segment that can be tracked together to allow the telescope to operate over an unvignetted elevation range of 23- 58 degrees. The gyro-stabilized telescope is moved by magnetic torquers around a 1.2-m diameter spherical hydrostatic bearing that floats under an oil pressure of 50 atmospheres within two closely fitting spherical rings mounted in the 21 foot diameter pressure bulkhead on the axis of the Nasmyth beam (see Figure 5). The focal plane instruments and the observers are on the pressurized side of bulkhead, allowing a room temperature working environment for the researchers and crew (Figures 6 and 7). The travel of the bearing for azimuth tracking is only +/- 3 degrees, and the airplane's heading must be periodically adjusted to keep the source under observation within this range. Each specific scientific observational program therefore requires its own customized flight plan which requires occasional adjustments by the telescope control system /flight software to compensate for departures from planned values of wind speed and direction.

The telescope optics will provide 1.6 arcsecond images on-axis (80% encircled energy) at 0.6μm; diffraction-limited performance will eventually be achieved at wavelengths longer than 15μm. Image quality of 2.8 arcseconds (FWHM, 5μm) using dynamical flexure compensation and an RMS pointing stability of 1.6 arcseconds were measured at first light (see Section 3), with future image quality anticipated to be ~ 1.1 arcseconds FWHM at this same wavelength with a pointing stability goal of less than 0.5 arcseconds RMS during the full operations phase that begins in 2014. A pointing accuracy of better than 0.5 arcsecond will be possible for observations where on-axis focal plane tracking can be used.

*2.2 SOFIA's first generation of instruments*

The names and operational characteristics of SOFIA's seven first generation Science Instruments (SIs) are summarized in Table 2. Together, they cover a much wider range of wavelengths and spectral resolutions than do the SIs of any other observatory (Figure 8). The SOFIA SIs include three Facility Class Science Instruments (FSIs) that will be maintained and operated by the SMO staff: the Faint Object InfraRed CAmera for the SOFIA Telescope (FORCAST), the High-resolution Airborne Wideband Camera (HAWC), and the First Light Infrared Test Experiment CAMera (FLITECAM). FSI pipeline-reduced and flux-calibrated data will be archived for general access by the astronomical community after a one year exclusive access (proprietary) period.

The remaining four SIs are Principal Investigator (PI) class instruments, maintained and operated by the PI teams at their home institutions. These instruments are designed to be less general in their potential applications than FSIs and are more likely to undergo upgrades between flight series so that they can be maintained at the state-of-the-art. General investigators (GIs) will be able to propose for PI instruments in collaboration with the PI team. Present development plans are for pipe-line reduced data from the US PI instruments to be added to the science archive after a one year exclusive access period. Two PI-class instruments are being developed in the US, Echelon-Cross-Echelle Spectrograph (EXES) and High-speed Imaging Photometer for Occulations (HIPO). Two PI-class instruments being developed in Germany are the German REceiver for Astronomy at Terahertz Frequencies (GREAT) and the Field Imaging Far-Infrared Line Spectrometer (FIFI LS). Current plans are to offer FIFI LS to the SOFIA science community as a Facility-like instrument. The FIFI LS data will be pipeline-reduced and flux-calibrated before they are placed in the data archive.

*2.3 The expected performance of SOFIA's first generation instruments*

SOFIA is designed to observe at wavelengths from 0.3 μm to 1.6 mm, and its seven first generation instruments will be capable of both photometric imaging and high resolution spectroscopy (Figure 8). The 8 arcminute diameter field of view (FOV) at the Nasmyth focus is compatible with the use of very large format detector arrays in the future. Although the thermal IR background at SOFIA's operational altitude is higher than the backgrounds seen by cryogenically-cooled space infrared telescopes, SOFIA's 2.5-m clear aperture telescope will still be capable of sensitivities (Figures 9, 10, and 11) about an order of magnitude better and linear spatial resolutions (Figure 12) more than five times better than those achieved by the Infrared Astronomical Satellite (IRAS, Neugebauer et al. 1984). SOFIA will match or exceed the sensitivity of the Infrared Space Observatory (ISO, Kessler et al. 1996) at high spectral resolution (Figure 9). In addition, the spectral resolution offered will be far higher than was available with Spitzer. SOFIA's spatial resolution will be about three times better relative to *Spitzer* at wavelengths longer than ~ 10 μm. Its capability for diffraction-limited imaging beyond 15 μm will enable SOFIA to make the sharpest images of any current or planned IR telescope operating in the 30 to 60 μm region (Figure 12). When Herschel's cryogens are depleted in 2013, SOFIA will be the only NASA mission with 25–600 l μm capabilities for many years. Important follow-up of Herschel (and Planck) discoveries can be made with SOFIA.

*2.4 Future Instrumentation development for SOFIA*

A tremendous advantage of an airborne observatory like SOFIA over space-based missions is its ability to make rapid and continuous instrumentation upgrades in response to new technological developments. Far-IR technology is advancing steadily, and major advances in sensitivity and array size are expected during the lifetime of SOFIA. The prospects were recently explored at two workshops. The first, entitled "SOFIA's 2020 Vision: Scientific and Technological Opportunities," was held at Caltech during December 6-8, 2007 (Young et al. 2008). The second, "Scientific Opportunities For new Instrumentation, Asilomar 2010," was held at Asilomar Conference Grounds on Monterey Peninsula June 6-8, 2010 (Asilomar 2010)

The SOFIA Project will support a technology development program that will regularly provide such opportunities. The next call for new instruments will be in mid-2011, and additional calls every 3 years will result in one new instrument or major upgrade to the observatory instrumentation every 18 months, on average. Approximately $5 to $10 M per year is the anticipated budget for instrument development funds for the life of the program. Possible future instrumentations capabilities projected during these workshops included:

- Expanded heterodyne wavelength coverage to enable more complete chemical studies of the ISM.

- Arrays of heterodyne detectors to increase observing speed and enable spectral mapping at high velocity resolution.

- Polarimeteric camera to probe the structure of magnetic fields in molecular clouds.

- Integral field unit spectrometer operating at R~2000 over the 5 to 60 μm spectral region to facilitate studies of the ISM on galactic scales.

- A variety of new spectrometers designed to facilitate detailed studies of the physics and chemistry of interstellar, circumstellar, and galactic environments.

## 3. Results of the Telescope Assembly Characterization and First Light Flight

SOFIA took to the air at 9:45 PM CDT on May 25, 2010 to conduct tests to characterize the performance of the telescope, as well as to acquire its first-light IR image. The first light images taken of the interacting starburst galaxy M82 (Figure 14) and Jupiter (Figure 15) were obtained early on May, 26, 2010 with the FORCAST Camera at an altitude of 35,000 feet.

*3.1 Planning and Objectives*

Planning the Telescope Assembly Characterization and First Light (TACFL) observations proved to be significantly more challenging than is expected for flights during nominal operations due to the significant number of constraints imposed on the operation of an aircraft and telescope under development. The aircraft had not yet completed test flights that would enable it to fly at altitudes exceeding 35,000 ft. In addition, because of its development status it could only fly in restricted airspace. The combination of these constraints motivated the decision to fly the observing legs over military-restricted airspace off the coast of southern California, However, confining the aircraft to this area imposed a maximum amount of observing time to less than 40 minutes per object. Additional constraints were associated with the telescope itself: due to the fact that the TACFL flight was scheduled before the end of the last aircraft test flight, the telescope was not permitted to execute slews during the flight, but rather had to remain fixed at an elevation of 23°. Finally, to protect the telescope from inadvertently catching sunlight in the event of an unexpected open-door landing (Figure 13), the flight plan had to schedule the aircraft to land before twilight.

The TACFL flight represented several "firsts" for the observatory: not only did the observations represent the first detection of infrared light through the system while in flight, but this flight was the first one performed at night with the cavity door open, for which scientists were onboard or that a science instrument was operated, and for which a carefully laid out flight plan coordinated the observations of specific objects. Therefore, more than just producing a first-light image, the TACFL event was an initial "out of the box" demonstration of the telescope's performance at this as-yet incomplete phase in the facility's development.

The top priority of the TACFL flight was to characterize the telescope's performance; a secondary priority was to acquire a media-worthy IR first light image. Specifically, the primary goals of the TACFL flight included exercising the chopper, determining how well the telescope could remain gyroscopically stabilized, investigating pointing control and sensitivity of focus to ambient temperature and to temperatures of various components within the telescope assembly. An initial assessment of at-altitude IR image quality as seen through the FORCAST detector was also an objective. Objectives for the TACFL flight extended beyond evaluating performance of the telescope itself. This milestone also provided an opportunity to exercise flight planning, to assess associated efficiencies (e.g., exploring ways to minimize the amount of time spent on "dead legs" in which the observatory is not collecting data), to determine how effectively the pilots could manually hold to the flight plan, and to observe the operation of the integrated flight crew. The flight crew, which was composed of several different teams (e.g., mission operations, telescope operators, science instrument team, pilots, etc), had never worked together on a flight before the TACFL flight.

*3.2 First Light Images of M82 and Jupiter*

Of the several constraints, described above that were imposed on the TACFL flight, the fixed elevation angle of the telescope, in combination with the limited length of flight legs, limited how long that objects could be tracked. Typically, the tracking time could be no more than ~ 40 minutes, imposing an upper limit on the target magnitudes. Jupiter and M82 were selected as First Light objects because they are IR-bright objects with extended emission that had significant potential for producing a visually-interesting image for press release. Also, such data for these objects were unique, as the objects had never before been observed at the selected bandpasses. Finally, these objects were also selected to symbolically represent the anticipated future contributions of SOFIA to a broad range of astrophysics, ranging from Solar System to extragalactic investigations.

Images of Jupiter were taken through filters centered at 5.4, 24.2 and 37.1 μm, while M82 was imaged at 19.7, 31.5 and 37.1 μm. Standard data reductions, including subtraction of 2 pairs of on-detector chopped images separated by an on-detector nod were recorded to eliminate background. The FWHM of the PSF in the 37.1 μm images was about 4 arcseconds, which is close to the diffraction limit. For each filter, a composite image was produced by stacking the 4 individual images. Typical integration time per frame was 30 seconds, resulting in signal-to-noise ratios exceeding 100. The final image of Jupiter was deconvolved with a point-spread function derived from Jupiter's moon Ganymede, which had an angular diameter of 1.37 arcseconds on the date of observation. Three-color images, presented in the first-light press release were produced for both Jupiter and M82 from these stacked images.

The FORCAST images taken of M82 (Figure 14), the archetype starburst galaxy, demonstrate the power of mid-infrared imaging: star-forming knots are clearly discerned through the dust lanes that obscure the galaxy nucleus at visible wavelengths. The observations of Jupiter (Figure 15) were also of particular scientific interest, given the recent disappearance of Jupiter's prominent dark-colored band just south of the equator. That band is also absent in the 5 and 37 μm images, but does appear at 24 μm as a brighter region against the planet disk. The prominent band still visible at optical wavelengths just north of the equator is present in all three IR bandpasses as a radiative source, but is most strong at 5 μm, for which the brightness temperature is over 200K. The 24.2 and 37.1 μm observations allow examination of variations in temperature and of the para- vs ortho-$H_2$ ratios as a function of depth in the Jovian atmosphere (for the latter, see Burgdorf et al. 2003) and any correlations in these quantities with changes of the visible cloud structure. Of particular interest is a possible restoration of the normally dark South Equatorial Belt that was anomalously bright during the first-light observations. Differences across the planetary disk will diagnose the extent of a different vertical wind regime in the region and help constrain the correlation between dynamics and cloud chemistry.

*3.3 Results of the Analysis of the Image Size at First Light*

Analysis of the light profiles of sources in 24.2 and 37.1 micron images taken indicates a jitter contribution to the image size of about 2.7 arcseconds (FWHM). This measurement excludes diffraction effects. At these long wavelengths, the shear layer flowing over the cavity is thought to be an insignificant component of the image size. Thus, this jitter measurement is primarily that of image motion. The images are somewhat elongated in the cross-elevation direction. The image motion in this cross-elevation direction was found to be primarily due to downwash wind loading at 90Hz on the secondary spiders. The elevation jitter is in the 1 to 10 Hz range, and due to find drive control residuals. An actual jitter contribution of 2.7 arcseconds (FWHM) is dramatically better than the jitter requirement for the initiation of science flights of FWHM $\leq$ 3.5 arcseconds.

Figures 16 and 17 show the 5.4 micron image quality produced by the SOFIA telescope during the TACFL flight. Figure 16 is the short term Point Spread Function (PSF) on a 2.5 millisecond time scale using the observations of the star $\gamma$ Cygni. It has a FWHM size of 2.4 by 1.9 arcsec. At the present time the observatory does not have this shift and add mode with FORCAST in place for outside observers. Figure 17 is the long term PSF obtained during an equivalent integration time of 7 seconds. It has a FWHM of 4.4 by 3.5 arcsec.

*3.4 Plans for Future Improvement of the System and Expected Performance*

The TACFL flight data indicate that SOFIA image quality exceeds requirements for entering the initial science observations, but the requirements are considerably more stringent for the nominal operations phase following mission development. A significant number of improvements to the observatory, followed by a series of verification and validation flights, are planned to insure that the image quality requirement for nominal operations is met by 2013 (three years following the initiation of science flights). This future requirement calls for the PSF, excluding the effects of shear flow, to not exceed ~ 1.1 arcseconds (FWHM) at 0.55 microns. The dominant sources of PSF broadening that will motivate observatory performance improvement activities in the future include image jitter and chopper performance.

3.4.1 *Image Jitter Reduction*

Image jitter is wavelength independent, but due to the dominance of the diffraction-limited image size, becomes less obvious at longer wavelengths. Allocating realistic values for telescope optical quality and expected shear layer effects at 0.55 microns in the image quality error budget, the jitter will need to be reduced to at least 0.9 arcseconds (FWHM) in order to meet the specification for overall image quality (excluding shear layer effects) of ~ 1.1 arcseconds. The effects of shear flow, a PSF-broadening agent expected to dominate at optical wavelengths, cannot be eliminated or reduced without considerable efforts in the future. However, a detailed understanding of its effects as a function of wavelength, and a comparison to models, will be provided through data collected in a series of characterization flights planned for 2011.

Two different systems will compensate for or damp out unintended motion of the telescope that causes image jitter. The flexible body compensation (FBC), which is already installed, uses the secondary mirror drive and fine drive motors to correct for image motions caused by telescope optical element vibrations. There is a known time lag of 22.5 ms, however, that needs to be reduced in order to make the system more responsive. Reducing this lag to 10 ms is a realistic goal. The use of a system of active mass dampers, strategically placed at key locations on the telescope structure itself, reduce the vibrations of the optics. With respect to jitter reduction, the mass dampers will be key to the future tuning of the observatory. Preliminary tests performed on the ground with temporary installations show that these devices have significant potential to damp out the dominant vibration modes seen in flight, such as a 90-Hz telescope spider mode that has been identified. While they do not directly impact jitter suppression, the gyros do help with overall pointing stability especially for twilight and daytime observations. A near-future goal will be to minimize and calibrate gyro drift.

3.4.2 *Chopper performance improvements*

As mentioned above, images taken during the TACFL flight were slightly elongated in the cross-elevation direction. The chopper was also in the cross-elevation direction. While some of this elongation could be the result from the chopper, recent tests show that the dominant source of image size is due to telescope vibrations caused by wind loading discussed above.

Another undesirable feature of the chopper is coma that results from the secondary mirror being in an off-axis orientation when in the "throw" or "OFF" position. The coma introduces a ~1.3 arcsecond (FWHM) blur for each arcminute of the chopper throw. Just as in the case with shear layer effects, nothing can be done to address this source of image degradation, but its effects will be monitored to assess the impact to science productivity. Asymmetric chopping, observing the target of interest in the on-axis position and chopping off-axis, has been demonstrated to be a viable observing strategy for the imaging of extended sources with chopper throws as large as 7 arcmin with FORCAST.

**4. Early Science with FORCAST and GREAT**

Two phases of Early Science observations with FORCAST and GREAT are underway. The first phase, a series of three flights each with FORCAST and GREAT (six flights total), began during the fall of 2010 with FORCAST and will continue in early 2011 with GREAT. These flights, limited in scope and called Early Short Science, call for collaboration between the SI PIs and three General Investigator (GI) teams that were selected as a result of a Call for Proposal in 2008. These teams are lead by Mark Morris of UCLA and Paul Harvey of University of Texas, who used the FORCAST Instrument to image the Orion Nebula and S140 respectively, and David Neufeld of Johns Hopkins University, who will be supporting the GREAT Team science program.

The second phase of Early Science, Basic Early Science, will begin in mid-2011 with a series of 15 flights (12 by US observers and 3 by German observers) using FORCAST and GREAT. The call for proposals for these flights (see http://www.sofia.usra.edu/Science/proposals/basic_science/index.html ) was issued on April 19, 2010 and the proposal ingestion was completed at 11:59 PM PDT on July 30, 2010. Selections were announced in fall of 2010 and flight planning is underway in early 2011. Limited flight participation will be allowed for the winning teams, but there will still be close collaboration between the winning PIs and the SI PIs.

*4.1 FORCAST as Configured for Early Science*

FORCAST is a facility class, mid-IR diffraction-limited camera with selectable filters for continuum imaging in two bands (5-25 μm and 25-40 μm). For the Early Science call, the bands available were 5.4, 6.4, 6.6, 7.7, 8.6, 11.1, 11.3, 19.7, 24.2 μm for the Short Wavelength Camera (SWC), and 31.5, 33.6, 34.8, 37.1 μm for the Long Wavelength Camera (LWC). FORCAST will eventually incorporate low resolution (R = 200-800) grism spectroscopy in the 4-8, 16-25 μm and/or 25-40 μm regions, but these grisms are not available for Early Science. FORCAST will provide the highest spatial resolution possible with SOFIA at wavelength longer than ~ 5 μm, enabling detailed imaging of proto-stellar environments, young star clusters, molecular clouds, and galaxies. Simultaneous high-sensitivity wide-field imaging can be performed in the two-channels using 256 x 256 Si:As and Si:Sb detector arrays which sample at 0.75 arcseconds/pixel giving a 3.2 arcmin x 3.2 arcmin field-of-view. FORCAST can be used in a single channel mode or in a dual channel mode that uses a dichroic to split the incident light towards the short and long wavelength arrays, allowing one of the 11.1, 11.3, 19.7 or 24.2 μm short wavelength filters to be used simultaneously with any of the long wavelength filters. Of course, any filter can be used in the single channel mode to maximize sensitivity. For small objects (less than half the size of the array in the horizontal or diagonal directions), chopping and nodding can be performed on the array to increase sensitivity and shorten the overhead time associated with beam switching.

*4.2 GREAT as Configured for Early Science*

GREAT, a PI-class instrument, will investigate a wide range of astronomical questions requiring the highest spectral resolution. These include observations of the 158μm fine-structure transition of ionized carbon [CII], which is the most important cooling line of the cold interstellar medium and is a sensitive tracer of the star forming activity of a galaxy. GREAT is a 2-channel heterodyne instrument that offers observations in three frequency bands with frequency resolution down to 45 kHz ($R \sim 4 \times 10^{7}$).

For GREAT, the Basic Science Program offers three different observing modes. The supported observing modes are 1) Position switching (PSW), 2) Beam switching (BSW), and 3) On-the-fly mapping (in PSW or BSW mode). The receiver bands L #1 and L #2 will be operated simultaneously. Both have instantaneous bandwidths ~ 720 MHz. Different frequencies can be set in each band. Both back ends (two for each band) record data simultaneously. The frequency ranges of the two receiver bands are L #1 = 1.25 THz - 1.5 THz (200 - 240 microns) and L #2 = 1.82 THz - 1.92 THz (156-158 microns). Two array-Acousto -Optical Spectrometers (AOS) with 1 MHz resolution (R ~ $10^6$) and two Chirp-Transform-Spectrometer (CTS) spectrometers: 220 MHz bandwidth and 47 kHz resolution. are used as backends.

*4.3 Science Projects Selected for Early Science*

A wide variety of investigations was selected for the SOFIA Early Science Program. The science areas covered included the Interstellar Medium (ISM), star formation, evolved stars, nearby galaxies, and planetary science. Twenty one proposals were awarded a total of 52.1 hours of FORCAST time and six proposals were awarded 17.4 hour of time with GREAT. The titles and abstracts of the selected proposals can be viewed at
http://www.sofia.usra.edu/Science/proposals/basic_science/accepted.html

**5.0 Summary**

The Stratospheric Observatory for IR Astronomy (SOFIA) will be the premier platform from which to make many imaging and spectroscopic astronomical observations in the 0.3 μm to 1.6 mm spectral region for the next twenty years. With its access to regions of the atmosphere that are opaque from the ground, its rapid and global deployability, and its ability to incorporate new and updated instruments, SOFIA will play an important role studying a variety of key astrophysical problems well into the first third of the 21st century.

**6.0 Acknowledgments**

We thank the entire SOFIA team for their tireless work on the SOFIA Project. R. D. G. was supported by USRA, NASA, and the US Air Force. Many thanks to Glenn Orton for assisting with the interpretation of the SOFIA first light observations of Jupiter.

Table 1: SOFIA system characteristics

| Characteristic | Value |
|---|---|
| Nominal Operational Wavelength | 0.3 to 1600 μm |
| Primary Mirror Diameter | 2.7-m |
| System clear aperture diameter | 2.5-m |
| Nominal system f-ratio | f/19.6 |
| Primary mirror f-ratio | f/1.28 |
| Telescope's unvignetted elevation range | 20 to 60 degrees |
| Unvignetted field-of-view | 8 arcmin |
| Image quality of telescope optics at 0.6 μm | 1.6 arcseconds on-axis (80% encircled energy) |
| Diffraction limited image size | $0.1 \times$ (λ in μm) FWHM in arcseconds |
| Diffraction Limited Wavelengths | $\geq 15$ μm |
| Optical Configuration | Bent Cassegrain with chopping secondary mirror and flat folding tertiary |
| Chopper frequencies | 1 to 20 Hz for 2-point square wave chop |
| Maximum chop throw on the sky | ± 5 arcmin (unvignetted) |
| Pointing stability | 2.0 arc sec RMS at first light; 0.5 arcseconds RMS during full operations phase |
| Pointing accuracy | 0.5 arcsecond with on-axis focal plane tracking; 3 arcseconds with on-axis fine-field tracking |
| Total emissivity of telescope (goal) | 0.15 at 10 μm with dichroic tertiary; 0.1 at 10 μm with aluminized tertiary |
| Recovery air temperature in cavity (and optics temperature) | 240K |

Table 2: SOFIA's First Generation Instrument Complex

| Instrument | Description | Institution and PI | λ range (μm) Resolution (λ/Δλ) | Field of View Array Size Array Type | Date Available |
|---|---|---|---|---|---|
| **FORCAST** (Facility SI) | **F**aint **O**bject Infra**R**ed **CA**mera for the **S**OFIA **T**elescope: *Facility Instrument - mid-IR camera and grism spectrometer* | Cornell University<br><br>T. Herter | 5 – 40<br><br>Grisms:<br>R ~ 200-800 | 3.2' x 3.2'<br>256 x 256 @ 0.75"<br>Si:As, Si:Sb | 2010 |
| **GREAT** | **G**erman **RE**ceiver for **A**stronomy at **T**erahertz Frequencies: *PI Instrument – heterodyne spectrometer* | MPIfR, KOSMA, DLR-WS<br><br>R. Güsten | 60-240<br><br>$R = 10^6 - 10^8$ | Diffraction Limited<br><br>Single pixel heterodyne | 2011 |
| **FIFI-LS** (Facility SI – like modes to be provided) | **F**ield **I**maging **F**ar-**I**nfrared **L**ine Spectrometer: *PI Instrument with facility-like capabilities – imaging grating spectrometer* | MPE, Garching<br><br>A. Poglitsch | 42 - 210<br><br>R = 1000 - 3750 | 30" x 30" (Blue)<br>60" x 60" (Red)<br>2 -16 x 5 x 5 Ga:Ge | 2013 |
| **HIPO** | **H**igh-speed **I**maging **P**hotometer for **O**ccultation: *Special PI Instrument – high speed imaging photometer* | Lowell Observatory<br><br>E. Dunham | 0.3 – 1.1<br>R = UBVRI; custom NB filters | 5.6' x 5.6'<br>1024 x 1024 @ 0.05" or 0.33"<br>CCD | 2012 |
| **FLITECAM** (Facility SI) | **F**irst **L**ight **I**nfrared **T**est **E**xperiment **CAM**era: *Facility Instrument – near-IR test camera and grism spectrometer* | UCLA<br><br>I. McLean | 1 – 5<br><br>R ~ 2000 | 8.2' x 8.2'<br>1024 x 1024 @ 0.48"<br>InSb | 2012 |
| **HAWC** (Facility SI) | **H**igh-resolution **A**irborne **W**ideband **C**amera: *Facility Instrument – far-IR bolometer camera* | University of Chicago<br><br>D. Harper | 50-240<br><br>R = 5 - 10 | Diffraction Limited<br><br>12 x 32 Bolometer | 2013 |
| **EXES** | Echelon-Cross-Echelle (**EXE**) **S**pectrograph: *PI Instrument – echelon spectrometer* | University of California Davis<br><br>M. Richter | 5 – 28<br><br>$R = 10^4, 10^5$, or 3000 | 5" to 90" slit<br>1024 x 1024 As:Si<br>1" – 4" slit width | 2013 |

## 8.0 Figure Legends

Figure 1. With the sliding door over its 17-ton infrared telescope wide open, NASA's Stratospheric Observatory for Infrared Astronomy (SOFIA) soars over California's snow-covered Southern Sierras on a test flight on April 14, 2010.

Figure 2. The typical atmospheric transmission at a SOFIA observing altitude of 45,000 feet as compared to the transmission on a good night at Mauna Kea (13,800 ft. MSL). From 1 to 1000 μm, the average transmission is ≥ 80% except in the center of absorption lines mostly due to telluric $H_2O$. Background image: IRAC false color image of the Sombrero Galaxy, courtesy of NASA/JPL-Caltech.

Figure 3. SOFIA's flight lifetime and time-frame will make it the premier facility for doing far-IR and sub-millimeter wave astronomy from 2010 until the mid 2030s. It will be the only facility available for wavelength coverage in the 28-1200 μm spectral region and for high resolution spectroscopy during most of that period. The SPICA mission has yet to be formally approved.

Figure 4. A cut-away view of the SOFIA Observatory. (SOFIA Project Website 2009)

Figure 5. The bent Cassegrain-Nasmyth optical configuration of the SOFIA 2.5-meter infrared telescope. (SOFIA Project Website 2009)

Figure 6. Main cabin looking aft toward the pressure bulkhead and telescope assembly showing the facility class FORCAST imager mounted at the Nasmyth focus and members of the FORCAST Team. Left to right: Joe Adams (Project Scientist), Luke Keller (co-I), Terry Herter (PI), George Gull (Lead Engineer), Chuck Henderson (Mechanical Engineer), and Justin Schoenwald (Software Engineer). (NASA photo)

Figure 7. Main cabin observing stations showing members of the DSI and USRA operations team at work during the TACFL flight. Left to right: Holger Jakob, Randy Grashuis, Andreas Reinacher, and Uli Lampater. (NASA photo)

Figure 8. A comparison of the SOFIA first generation Science Instruments in terms of wavelength coverage and spectral resolution.

Figure 9. A comparison of the photometric sensitivity of several space and airborne infrared missions. SOFIA's photometric sensitivity will be comparable to that of ISO in the far infrared.

Figure 10. The continuum sensitivity expected for SOFIA's first generation science instruments at the onset of full operations. The minimum detectable continuum point source flux density (MDCF) in Janskys for a 10σ detection in 900 seconds of integration time are plotted against wavelength in μm. Observing and chopper efficiency are not included.

Figure 11. The line sensitivity expected for SOFIA's first generation spectrometers at the onset of full operations. The minimum detectable line flux (MDLF) in Watts per square meter for a 10σ detection in 900 seconds of integration time are plotted against wavelength in μm. Observing and chopper efficiency are not included.

Figure 12. SOFIA will form images three times smaller than those formed by the *Spitzer* Space Telescope and will have spatial resolution comparable to that of Herschel at wavelengths longer than 60 μm. (SOFIA Project Website 2009)

Figure 13. The SOFIA 747 SP aircraft was put through an extensive series of test flights to evaluate performance over the entire operational envelope. Here, with the door to the cavity housing its 2.5-meter infrared telescope wide open in a scheduled test to demonstrate that emergency door-open landings can be accomplished without dangerous control issues and/or contamination of the telescope cavity, *Clipper Lindberg* flares for landing at Air Force Plant 42 in Palmdale, Calif., after a test flight on April 14, 2010. (NASA Photo / Tony Landis)

Figure 14. Composite infrared image of the central portion of galaxy M82, from SOFIA's First Light flight, at wavelengths of 20 (blue), 32 (green) and 37 microns (red). The angular resolution is about 4 arcsec FWHM at all three wavelengths. The middle inset image shows the same portion of the galaxy at visual wavelengths. The infrared image sees past the stars and dust clouds apparent in the visible-wavelength image into the star-forming heart of the galaxy. The long dimension of the inset boxes is about 5400 light years. May 26, 2010 (Visual image by N. A. Sharp/NOAO/AURA/NSF)

Figure 15. Composite infrared image of Jupiter from SOFIA's first light flight at wavelengths of 5.4 (blue), 24 (green) and 37 μm (red), made by Cornell University's FORCAST camera during the SOFIA observatory's "first light" flight. The white stripe in the infrared image is a region of relatively transparent clouds through which the warm interior of Jupiter can be seen.

Figure 16. Top panel shows the "Short term" Point Spread Function (PSF) at 5.4 μm generated by co-adding 2,800 2.5 millisecond exposures of γ Cygni. The frames were registered at the pixel level before co-adding. The resultant PSF obtained in this way freezes out telescope jitter and to a lesser extent cavity seeing, in a manner analogous to the way the using the speckle imaging technique on a ground based telescope freezes out atmospheric seeing (Fried 1966). Bottom panel shows the lower (dashed line) and upper bounds to the image intensity profile obtained by fitting and an elliptical Gaussian curve to the 2.4 arcsecond x 1.9 arcsecond PSF.

Figure 17. Top panel (**top left**) shows the "long term" Point Spread Function (PSF) at 5.4 μm generated by the straight co-addition of 2,800 2.5 millisecond exposures (**top right**) of γ Cygni without registering the frames before co-adding. The resultant PSF obtained in this is the equivalent of a seven second integration that does not freeze out cavity seeing and telescope jitter. Bottom panel shows the lower (dashed line) and upper bounds to the image intensity profile obtained by fitting and an elliptical Gaussian curve to the 4.4 arcsecond x 3.5 arcsecond PSF. The image size at all FORCAST wavelengths in long exposures was approximately the same because it was limited primarily by image jitter.

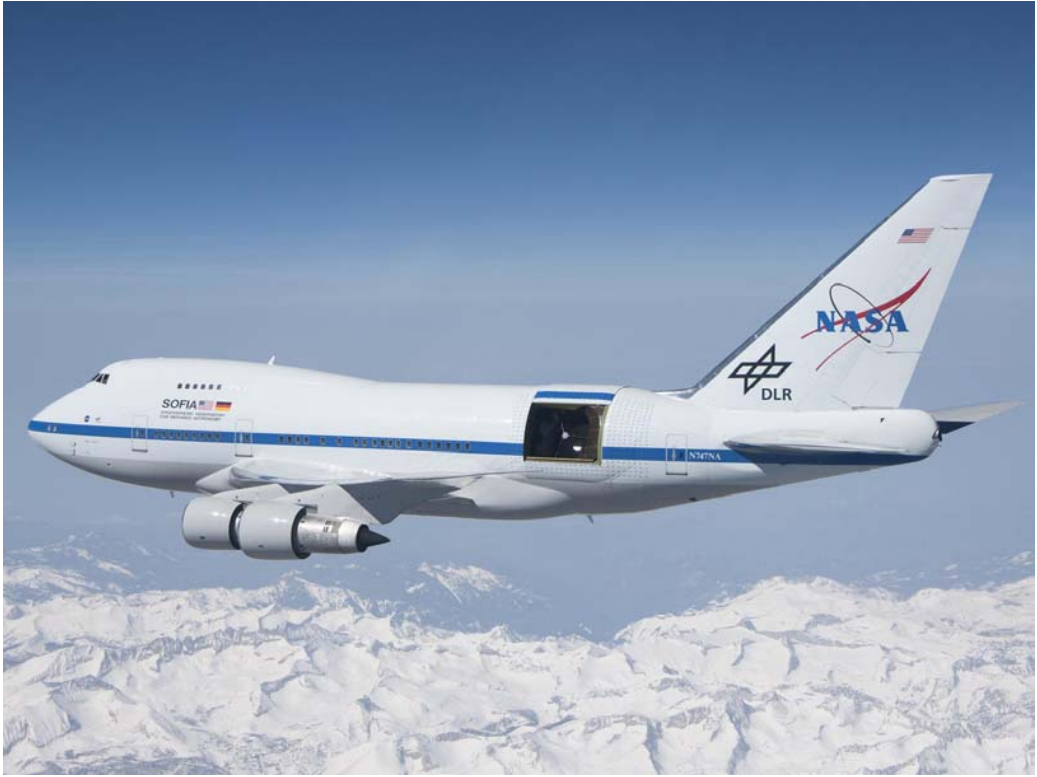

Figure 1. With the sliding door over its 17-ton infrared telescope wide open, NASA's Stratospheric Observatory for Infrared Astronomy (SOFIA) soars over California's snow-covered Southern Sierras on a test flight on April 14, 2010.  (NASA Photo)

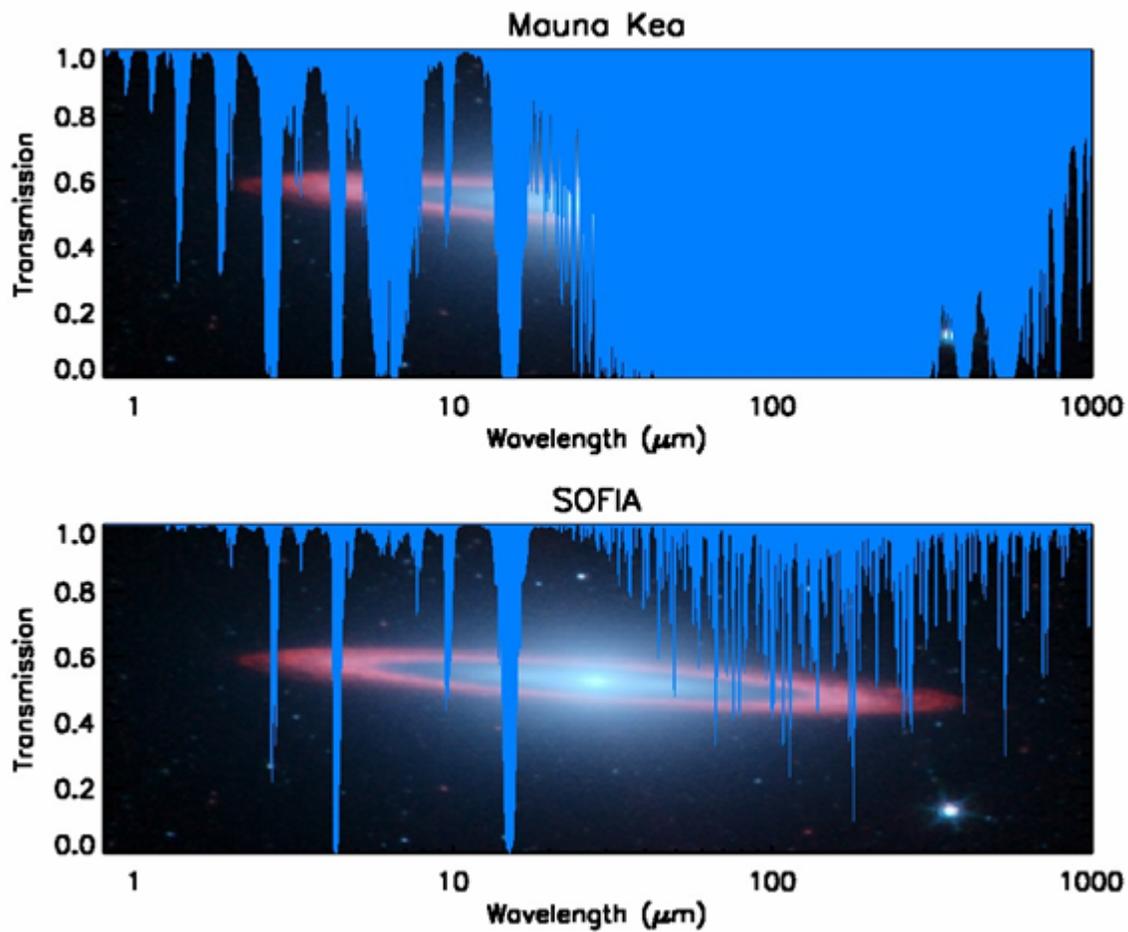

Figure 2. The typical atmospheric transmission at a SOFIA observing altitude of 45,000 feet as compared to the transmission on a good night at Mauna Kea (13,800 ft. MSL). From 1 to 1000 μm, the average transmission is ≥ 80% except in the center of absorption lines mostly due to telluric $H_2O$. Background image: IRAC false color image of the Sombrero Galaxy, courtesy of NASA/JPL-Caltech.

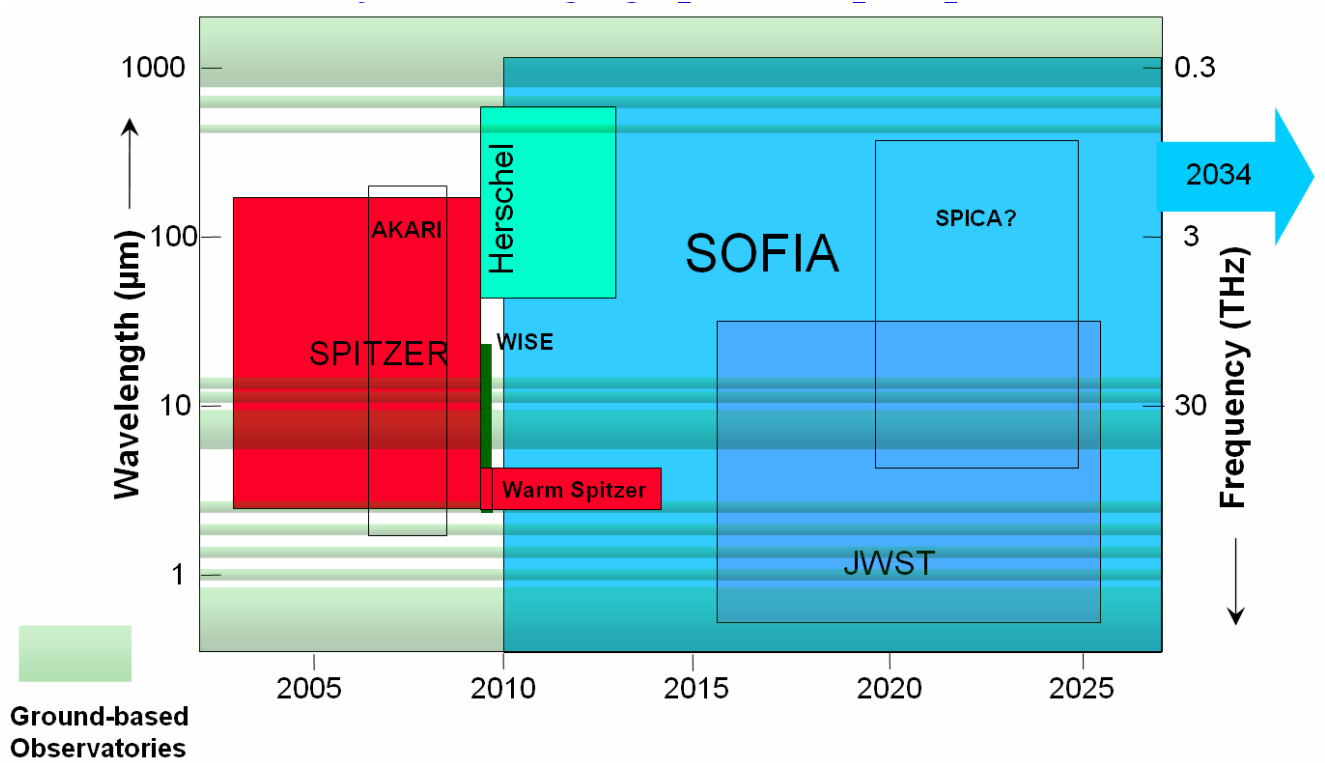

Figure 3. SOFIA's flight lifetime and time-frame will make it the premier facility for doing far-IR and sub-millimeter wave astronomy from 2010 until the mid 2030s. It will be the only facility available for wavelength coverage in the 28-1200 μm spectral region and for high resolution spectroscopy during most of that period. The SPICA mission has yet to be formally approved.

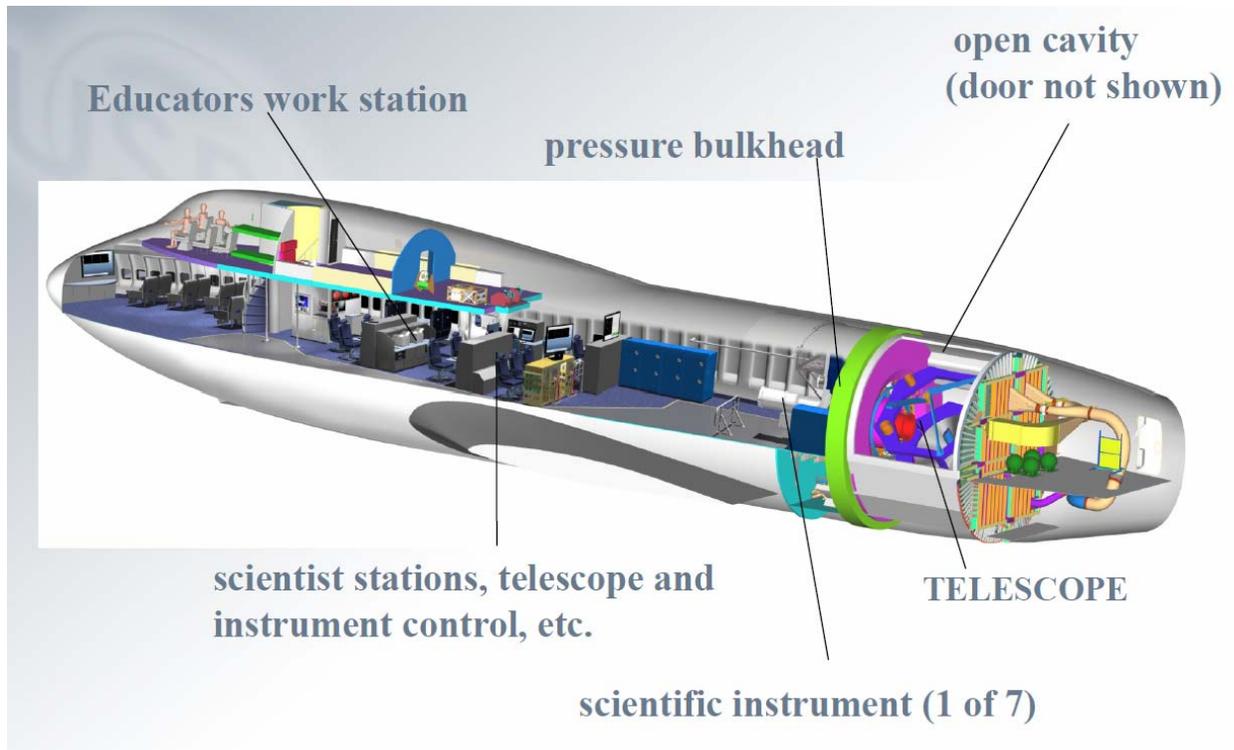

Figure 4.  A cut-away view of the SOFIA Observatory.  (SOFIA Project Website 2009)

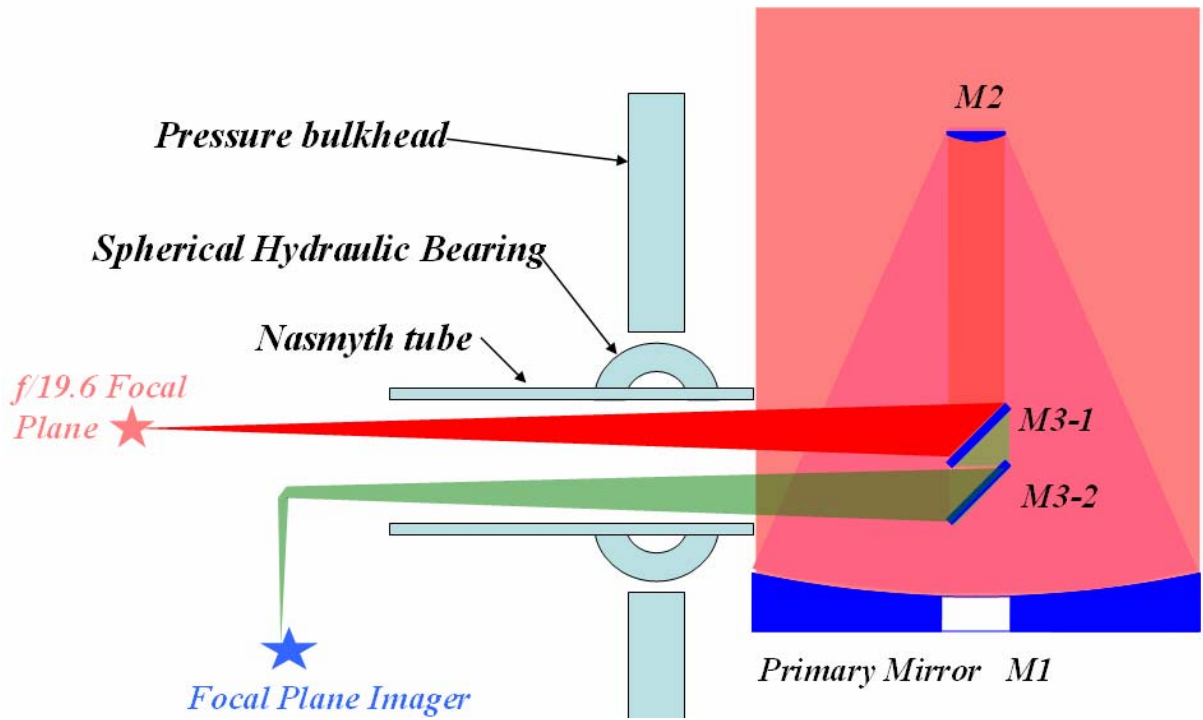

Figure 5. The bent Cassegrain-Nasmyth optical configuration of the SOFIA 2.5-meter infrared telescope. (SOFIA Project Website 2009)

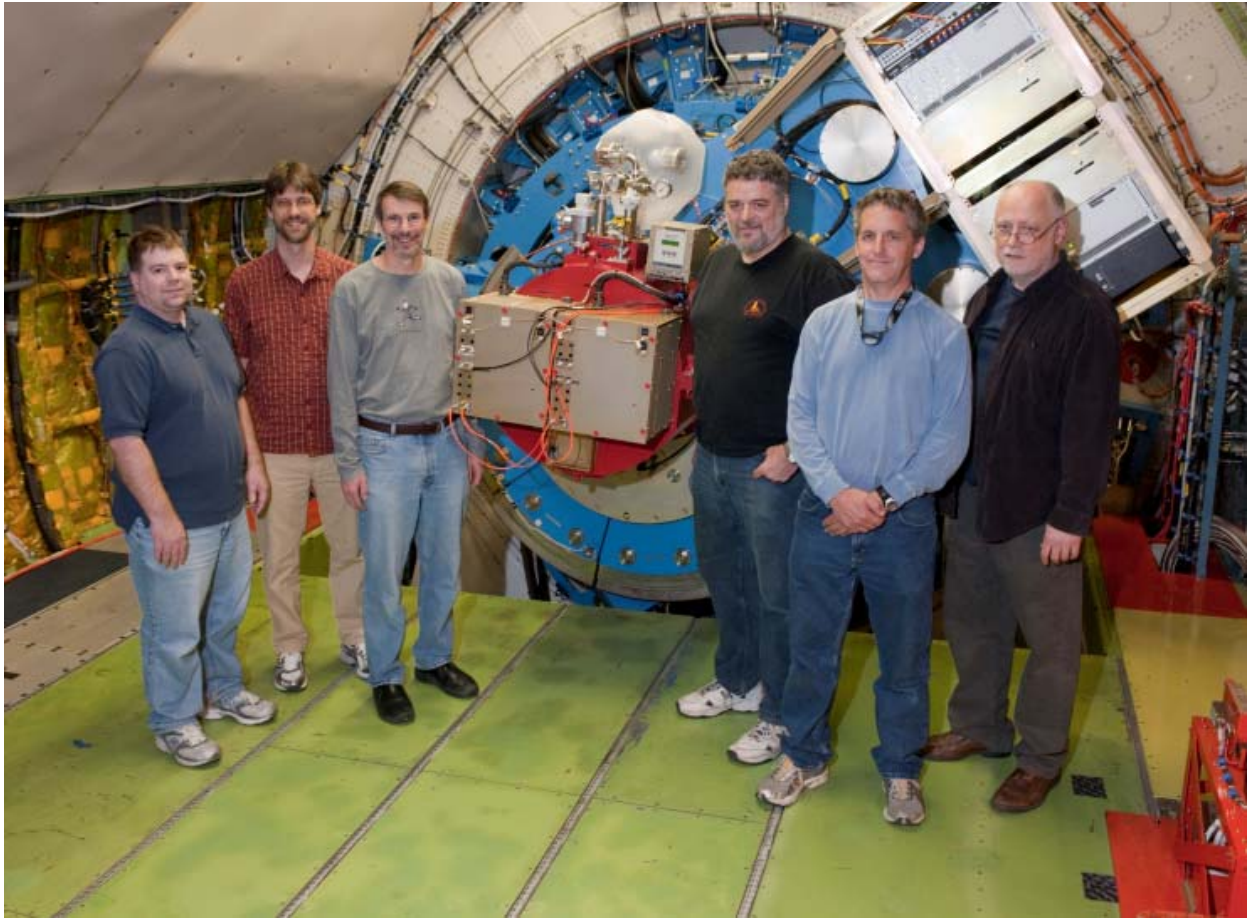

Figure 6. Main cabin looking aft toward the pressure bulkhead and telescope assembly showing the facility class FORCAST imager mounted at the Nasmyth focus and members of the FORCAST Team. Left to right: Joe Adams (Project Scientist), Luke Keller (co-I), Terry Herter (PI), George Gull (Lead Engineer), Chuck Henderson (Mechanical Engineer), and Justin Schoenwald (Software Engineer). (NASA photo)

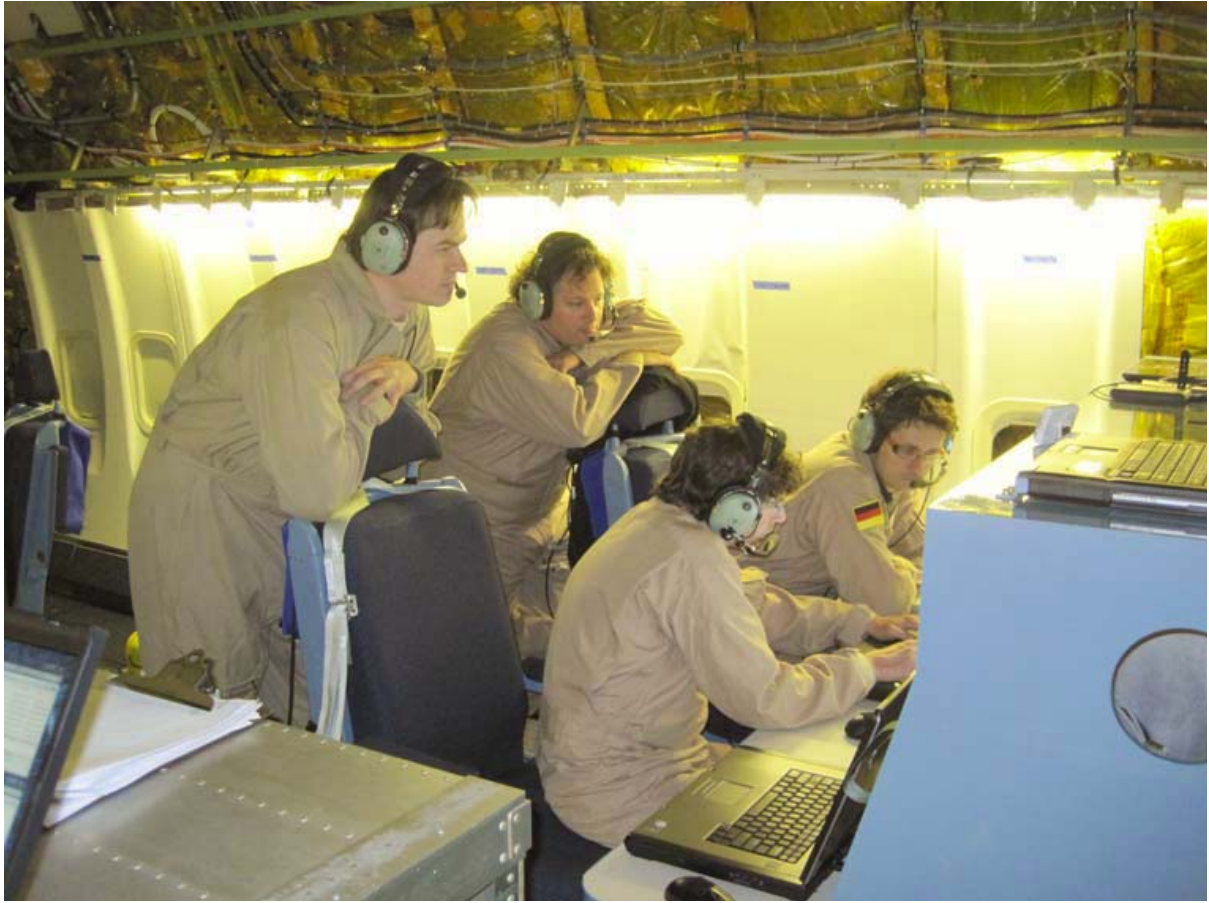

Figure 7. Main cabin observing stations showing members of the DSI and USRA operations team at work during the TACFL flight.   Left to right: Holger Jakob, Randy Grashuis, Andreas Reinacher, and Uli Lampater.  (NASA photo)

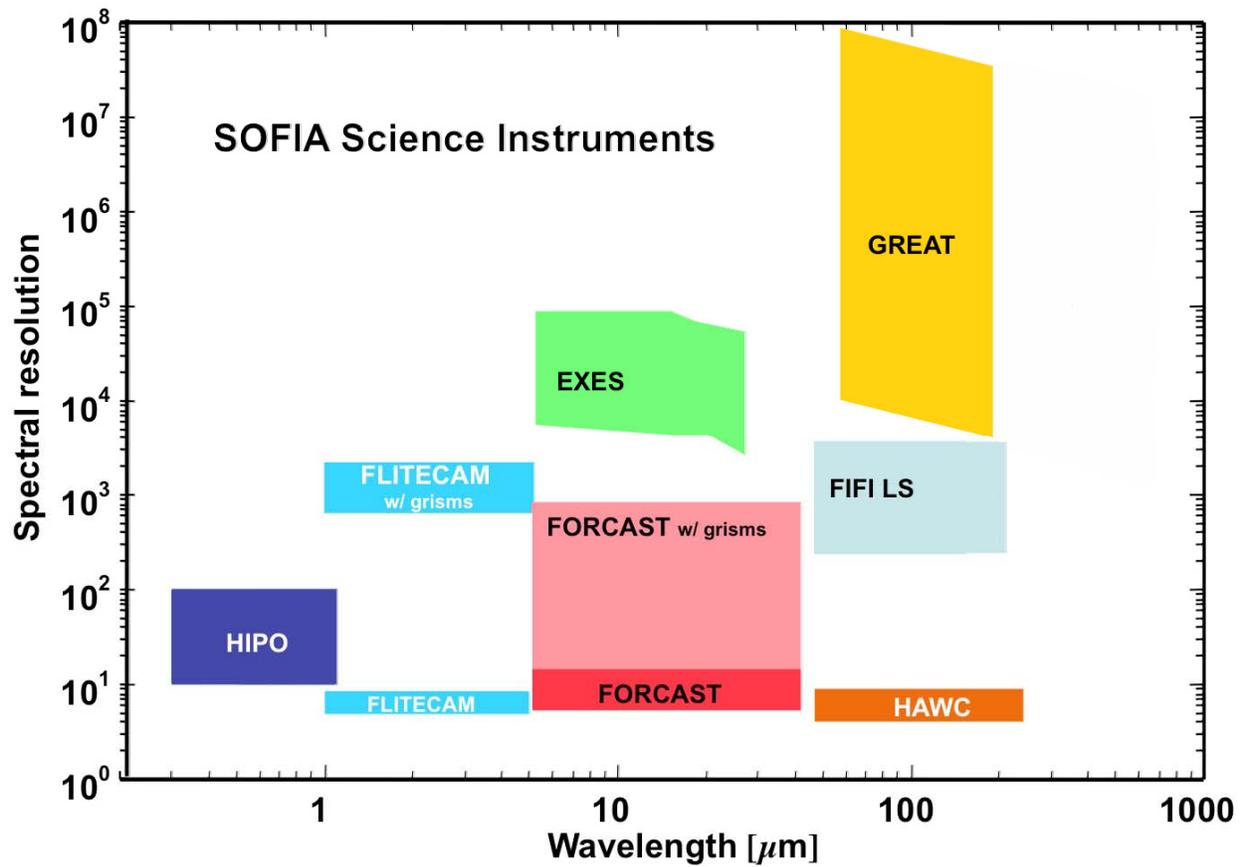

Figure 8. A comparison of the SOFIA first generation Science Instruments in terms of wavelength coverage and spectral resolution.

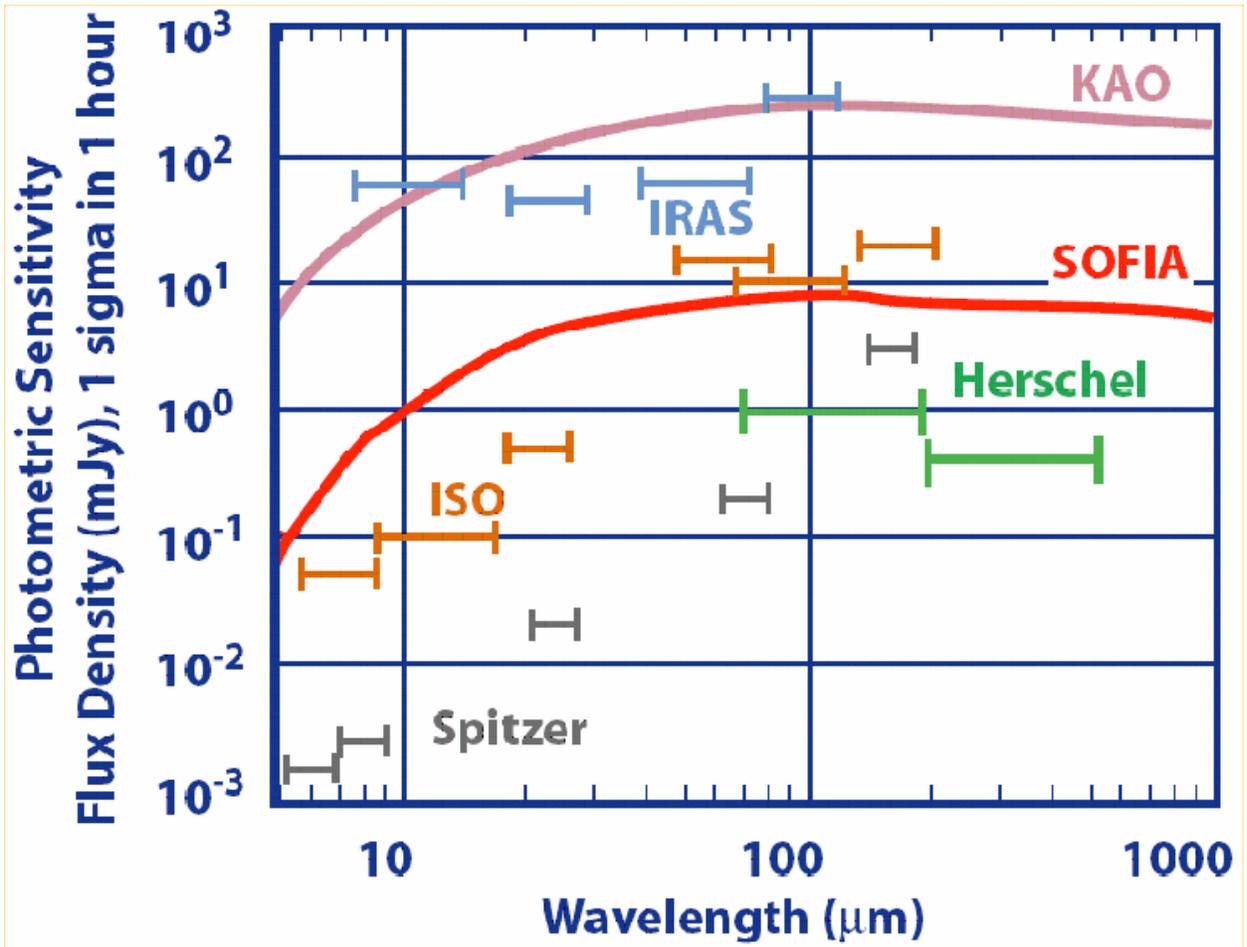

Figure 9. A comparison of the photometric sensitivity of several space and airborne infrared missions. SOFIA's photometric sensitivity will be comparable to that of ISO at far-infrared wavelengths. (SOFIA Project Website 2009)

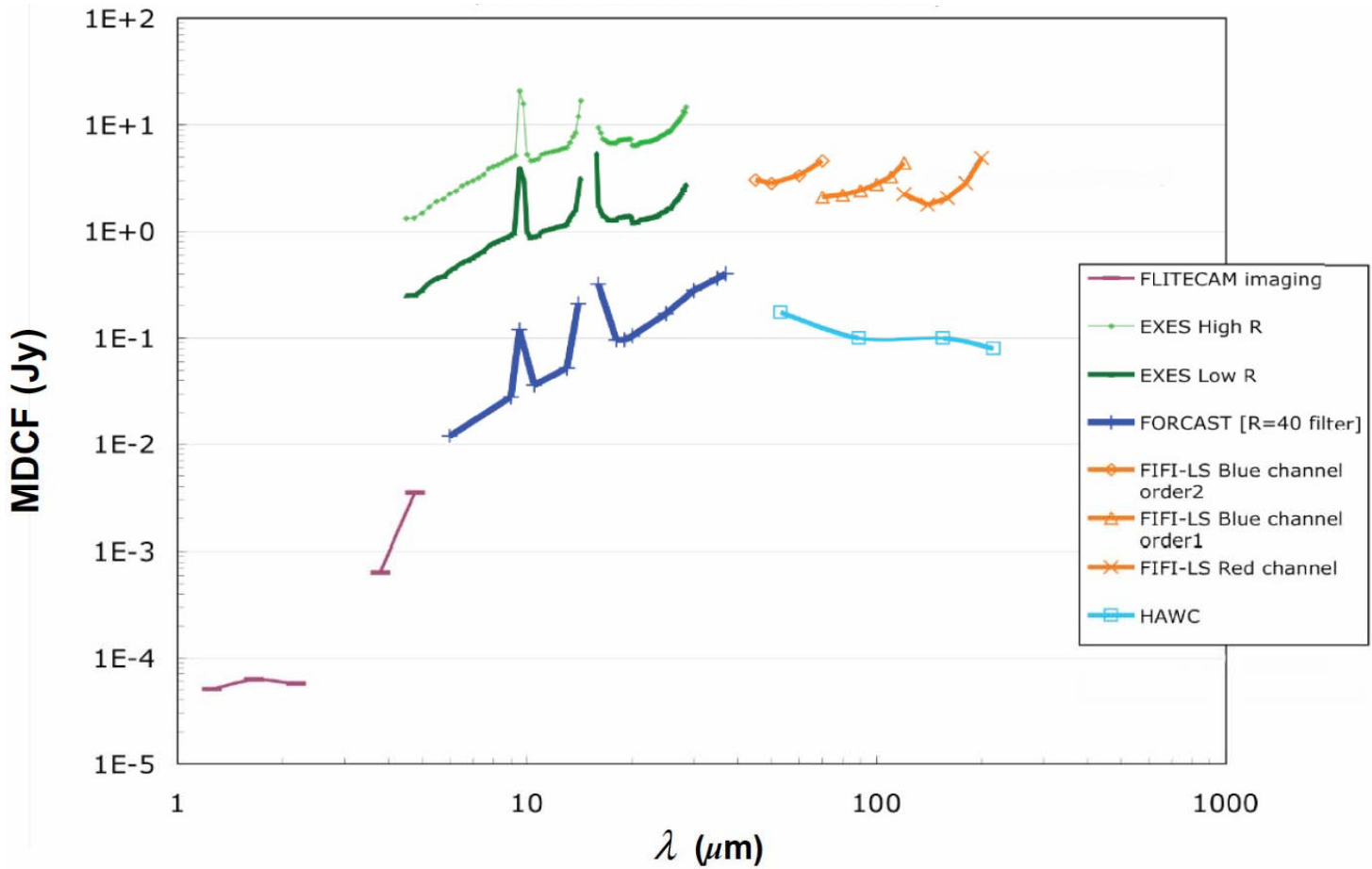

Figure 10. The continuum sensitivity expected for SOFIA's first generation science instruments at the onset of full operations. The minimum detectable continuum point source flux density (MDCF) in Janskys for a 10σ detection in 900 seconds of integration time are plotted against wavelength in μm. Observing and chopper efficiency are not included.

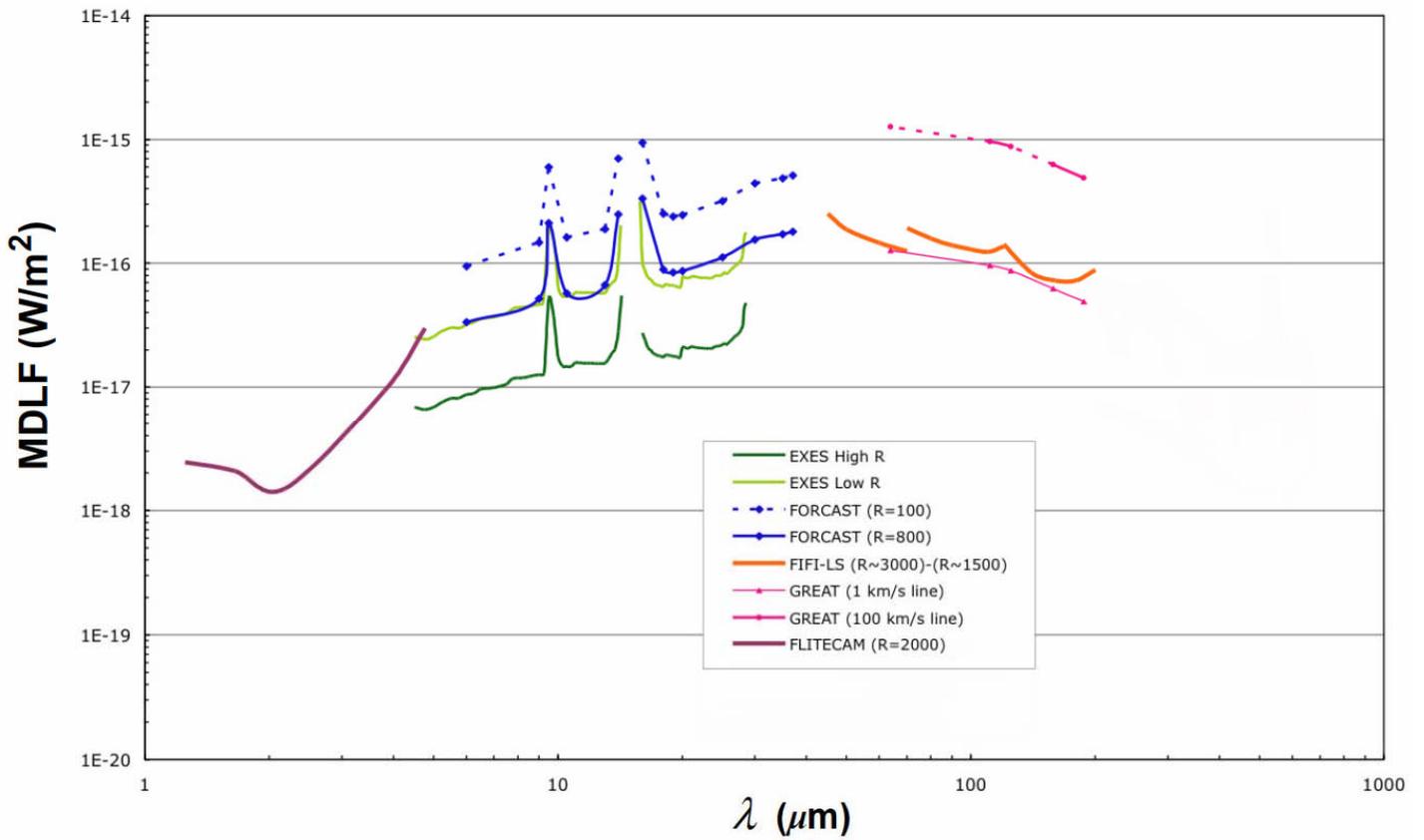

Figure 11. The line sensitivity expected for SOFIA's first generation spectrometers at the onset of full operations. The minimum detectable line flux (MDLF) in Watts per square meter for a $10\sigma$ detection in 900 seconds of integration time are plotted against wavelength in μm. Observing and chopper efficiency are not included.

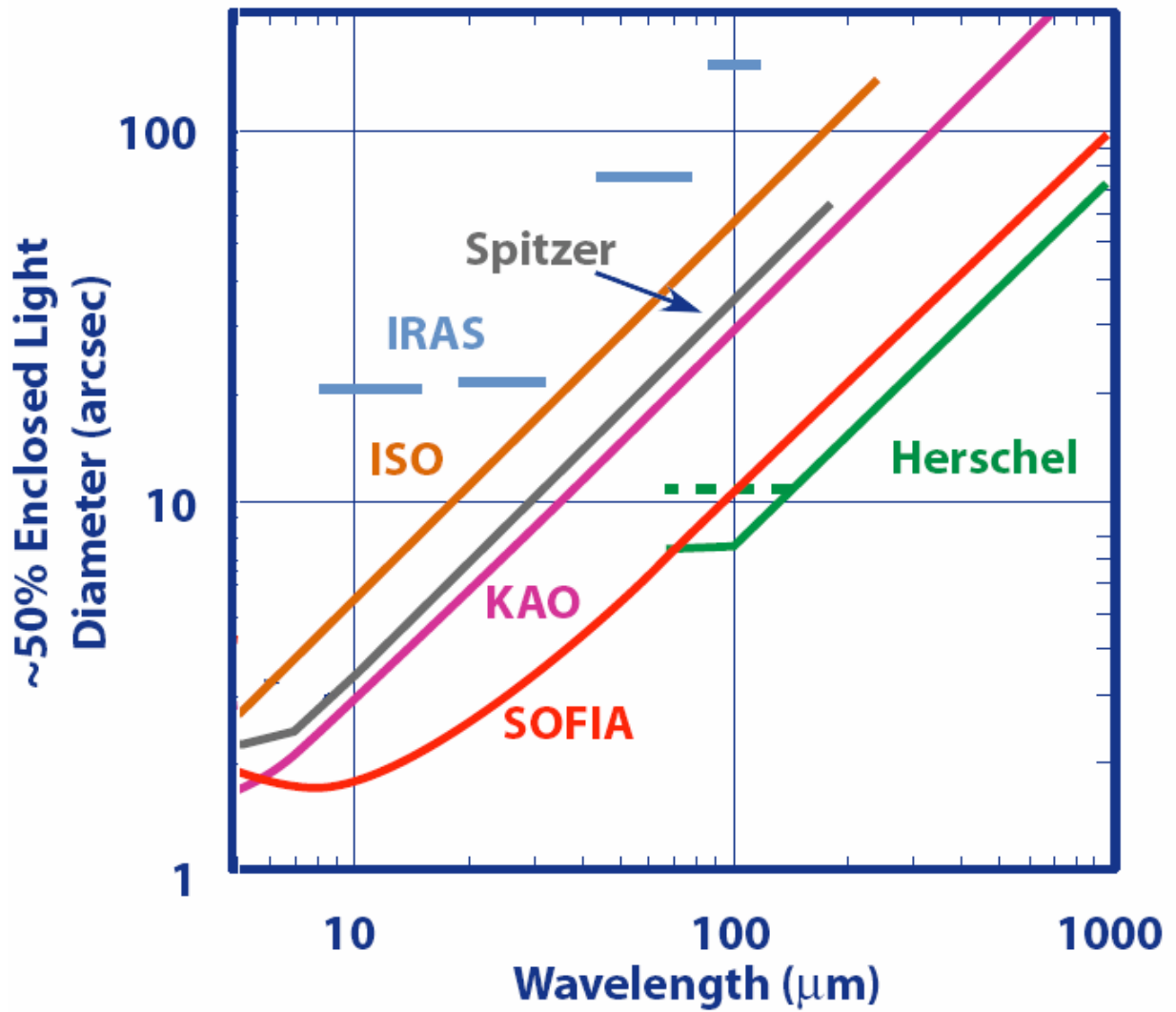

Figure 12. SOFIA will typically have PSFs three times smaller than those formed by the *Spitzer* Space Telescope and will have spatial resolution comparable to that of Herschel at wavelengths longer than 60 μm. (SOFIA Project Website 2009)

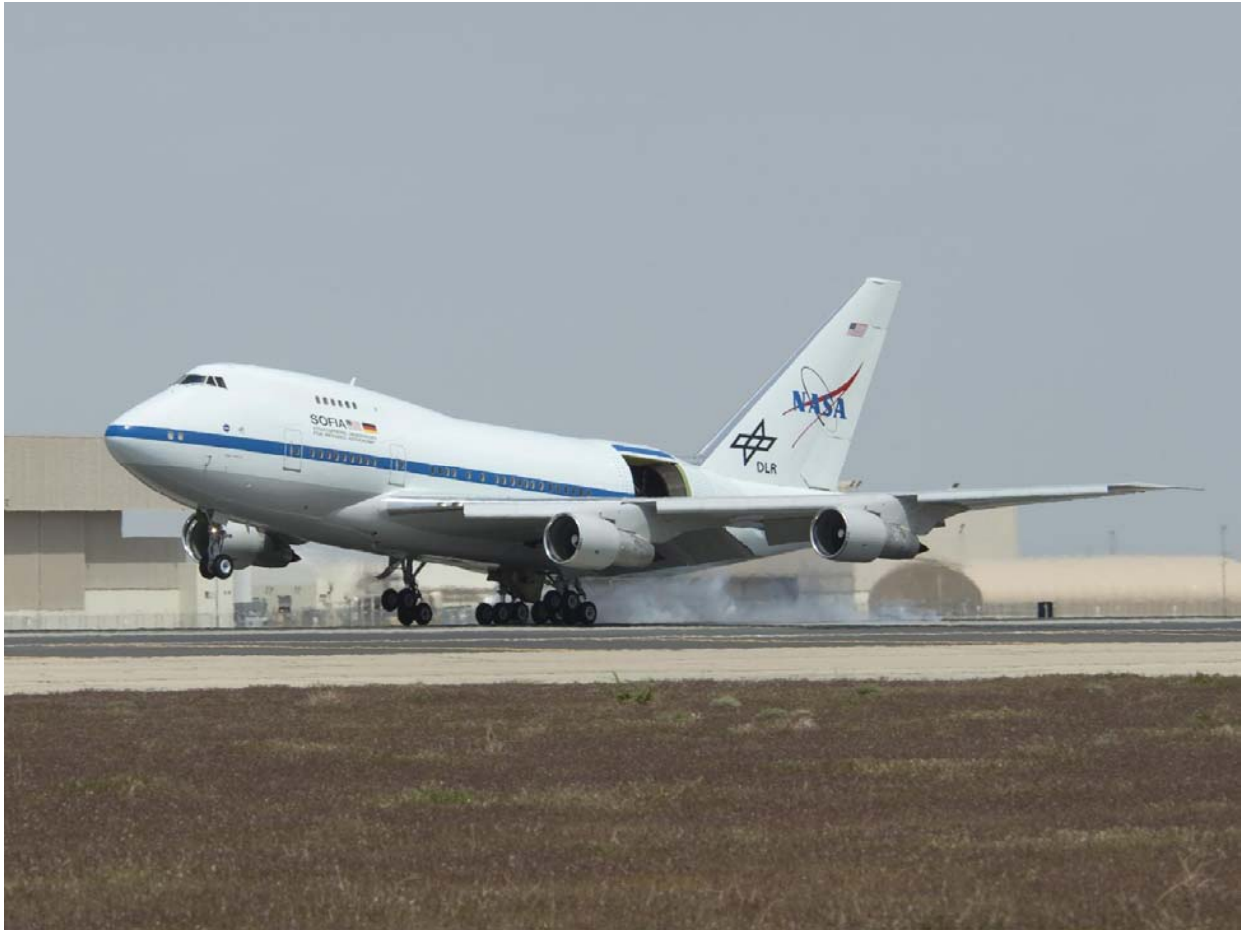

Figure 13. The SOFIA 747 SP aircraft was put through an extensive series of test flights to evaluate performance over the entire operational envelope. Here, with the door to the cavity housing its 2.5-meter infrared telescope wide open in a scheduled test to demonstrate that emergency door-open landings can be accomplished without dangerous control issues and/or contamination of the telescope cavity, *Clipper Lindberg* flares for landing at Air Force Plant 42 in Palmdale, Calif., after a test flight on April 14, 2010. (NASA Photo / Tony Landis)

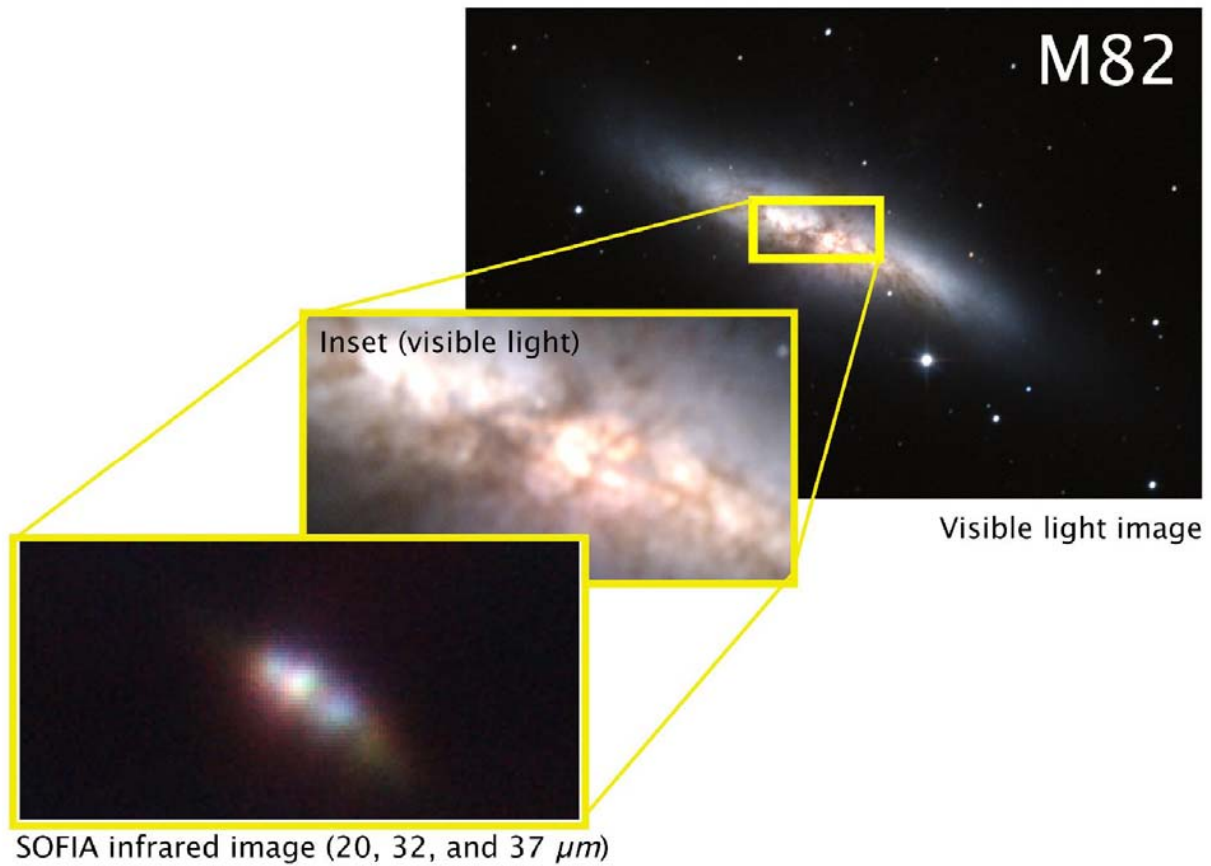

Figure 14. Composite infrared image of the central portion of galaxy M82, from SOFIA's First Light flight, at wavelengths of 20 (blue), 32 (green) and 37 microns (red). The middle inset image shows the same portion of the galaxy at visual wavelengths. The infrared image sees past the stars and dust clouds apparent in the visible-wavelength image into the star-forming heart of the galaxy. The long dimension of the inset boxes is about 5400 light years. Data taken on May 26, 2010. (Optical image by N. A. Sharp/NOAO/AURA/NSF)

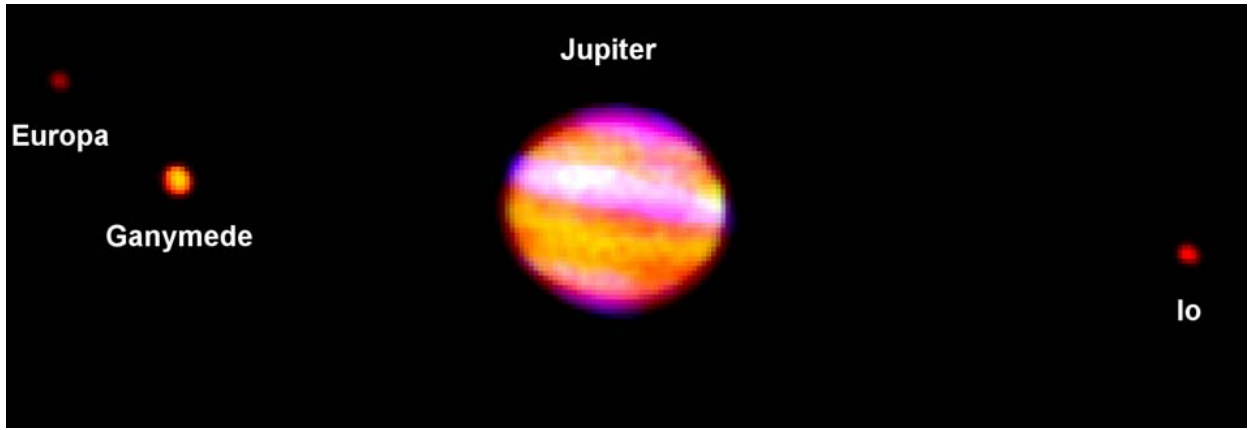

Figure 15. Composite infrared image of Jupiter from SOFIA's first light flight at wavelengths of 5.4 (blue), 24 (green) and 37 microns (red), made by Cornell University's FORCAST camera during the SOFIA observatory's "first light" flight. The white stripe in the infrared image is a region of relatively transparent clouds through which the warm interior of Jupiter can be seen. The images of three Galilean satellites, also observed with FORCAST during the TACFL flight, have been placed in their relative positions at the time of the observations. Data taken May 26, 2010.

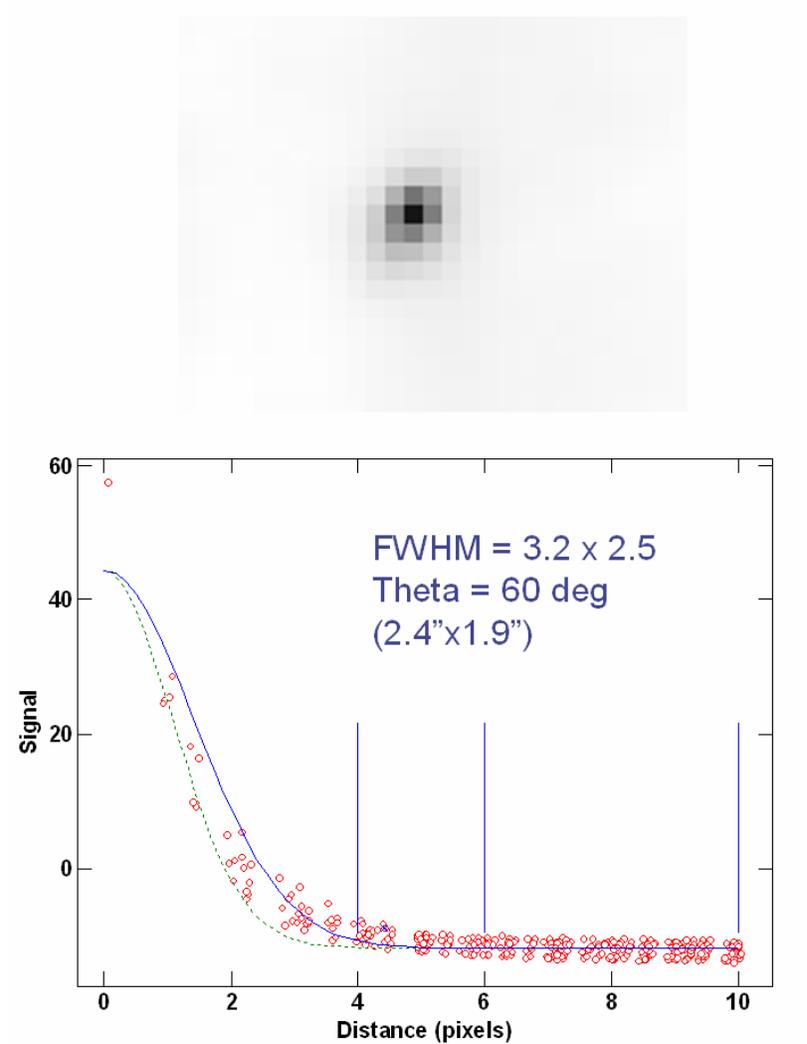

Figure 16. Top panel shows the "Short term" Point Spread Function (PSF) at 5.4 μm generated by co-adding 2,800 2.5 millisecond exposures of γ Cygni. The frames were registered at the pixel level before co-adding. The resultant PSF obtained in this way freezes out the telescope jitter and to a lesser extent the cavity seeing in a manner analogous to the way the using the speckle imaging technique on a ground based telescope freezes out atmospheric seeing (Fried 1966). Bottom panel shows the lower (dashed line) and upper bounds to the image intensity

profile obtained by fitting an elliptical Gaussian curve to the 2.4 arcsecond x 1.9 arcsecond PSF. The position angle of the major axis of the ellipse is 60°.

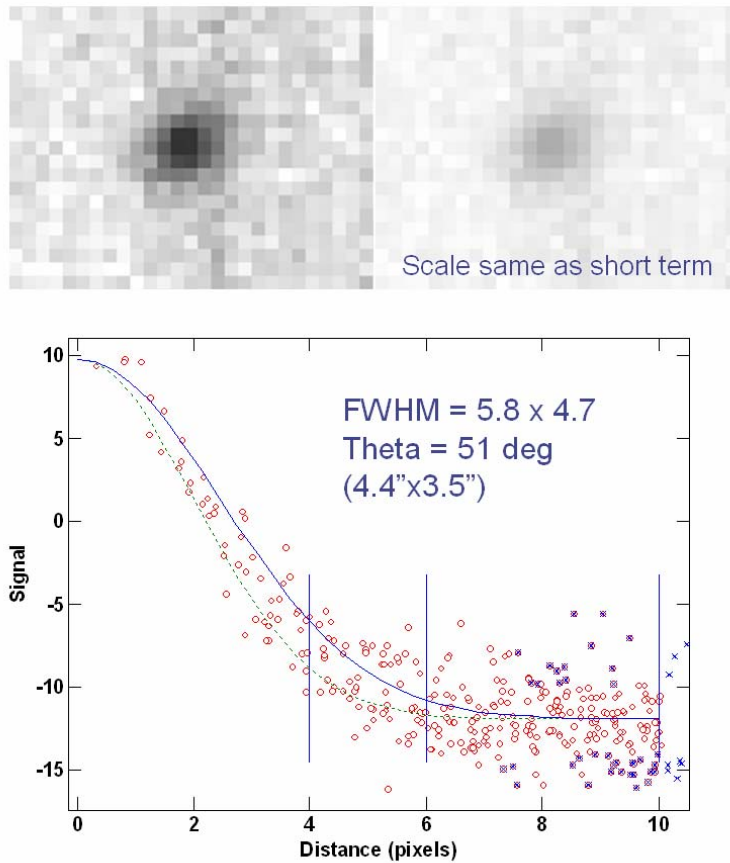

Figure 17. Top panel (**top left**) shows the "long term" Point Spread Function (PSF) at 5.4 μm generated by the straight co-addition of 2,800 2.5 millisecond exposures (**top right**) of γ Cygni without registering the frames before co-adding. The resultant PSF obtained in this is the equivalent of a 7 second integration that does not freeze out cavity seeing and telescope jitter. Bottom panel shows the lower (dashed line) and upper bounds to the image intensity profile obtained by fitting an elliptical Gaussian curve to the 4.4 arcsecond x 3.5 arcsecond PSF. The position angle of the major axis of the ellipse is 51°. The image size at all FORCAST

wavelengths in long exposures was approximately the same because it was limited primarily by image jitter.